# Multiple superconducting phases in a nearly ferromagnetic system.

D. Braithwaite[1], M. Valiska[1], G. Knebel[1], G. Lapertot[1], J.-P. Brison[1], A. Pourret[1], M.E. Zhitomirsky[1], J. Flouquet[1], F. Honda[2], D. Aoki[2].

[1]Univ. Grenoble Alpes, CEA, IRIG, PHELIQS, 38000 Grenoble, France.

[2]IMR, Tohoku University, Oarai, Ibaraki, 311-1313, Japan

abstract

Multiple superconducting order parameters are extremely rare. Here we show that a very small pressure can induce this phenomenon in the recently discovered heavy fermion superconductor $UTe_2$. This nearly ferromagnetic system shows several intriguing phenomena, including an extraordinary reinforcement of superconductivity in very strong magnetic fields. We find that pressure can tune the system to a more correlated state and probable magnetic order. The superconducting critical temperature is strongly enhanced, reaching almost 3K, a new record for Ce- and U-based heavy fermion superconductors. Most spectacularly under pressure we find a transition within the superconducting state, putting $UTe_2$ among the very rare systems having multiple superconducting phases. $UTe_2$ under pressure is a treasure trove of several of the most fascinating phenomena in unconventional superconductivity and may well be a keystone in their understanding.

In most superconductors the superconducting order parameter is s-wave, meaning it has the same symmetry as the crystal lattice. However in the ever expanding family of unconventional superconductors, which includes such disparate members as high-Tc cuprates and pnictides, organic superconductors, and heavy fermions, the order parameter can assume a number of different symmetries, usually lower than the lattice symmetry. This opens the intriguing possibility that a given system could in principle exhibit different order parameters, each one being selected by changing an external variable like temperature or magnetic field. This scenario does in fact exist, but is extremely rare, having been really established only in superfluid $^3$He[1] and in two superconductors: $UPt_3$[2,3] and thorium-doped $UBe_{13}$[4]. The recently discovered superconductivity in the heavy fermion system $UTe_2$[5] shows several unusual properties, the most spectacular being re-entrant superconductivity when magnetic fields as high as 60 Tesla are applied in specific directions[6-9]. Another intriguing property is the temperature dependence of the specific heat. Indeed in all samples a large residual term, of about 50% of the normal state specific heat, seems to remain as the temperature approaches zero[5,6]. In this report we show that the superconductivity is extremely sensitive to hydrostatic pressure as a tuning parameter and that $UTe_2$ is probably another example of multiple superconducting phases. The superconducting state found at zero pressure is monotonously depressed with pressure but a second superconducting state is found to emerge as pressure is increased. Pressure increases the splitting between the 2 transitions and the high temperature superconducting transition reaches nearly 3K, a new record for a U-based heavy fermion superconductor. Pressure also drives the system towards a more correlated state, with evidence for a strong enhancement of the electronic effective mass. At a critical pressure of about 1.7 GPa both superconducting states are suppressed and a new order parameter, probably magnetic, is found.

UTe$_2$ is a paramagnetic heavy fermion system, although its ground state seems to lie close to ferromagnetism[5,6]. The crystal structure is orthorhombic with an extremely anisotropic magnetic susceptibility at low temperatures, and a pronounced easy axis along the $a$-axis. When a magnetic field is applied along the hard $b$-axis, UTe$_2$ exhibits a metamagnetic transition at about 35T[7,10] and most spectacularly shows field enhanced superconductivity up to this field[8] and re-entrant superconductivity at even higher fields for specific angles of the applied field[9]. In this it shares some characteristics of the known ferromagnetic superconductors URhGe and UCoGe, which exhibit metamagnetism and enhanced or re-entrant superconductivity when a magnetic field is applied perpendicular to the easy axis[11,12]. However in these systems the enhancement of superconductivity is related to the suppression of ferromagnetism with a transverse field so the case of UTe$_2$ is quite different. Field enhanced superconductivity in a system without ferromagnetism offers further conceptual challenges and may well shed new light on this fascinating phenomenon. The remarkable properties and especially the superconductivity of these systems are undoubtedly linked to the proximity of a magnetic instability, and pressure is a powerful parameter to tune this proximity. Pressure can either push the system away from the instability, or more interestingly move the system closer to, and even cross the magnetic instability. In the vast majority of cases, superconductivity under pressure is studied by resistivity measurements. Zero resistance is the best-known characteristic of a superconductor as well as being easy to measure. But it is not necessarily the best probe, as zero resistance can be obtained with filamentary superconductivity or superconducting impurities and it is not proof of bulk superconductivity. Furthermore, once the resistance is zero, resistivity measurements are essentially blind to any further change of state that may take place inside the superconducting state. For this reason, the essential part of this study is obtained from calorimetry measurements, performed under pressure in a diamond anvil cell (See methods). This technique, although not quantitative, gives information that is directly related to the specific heat of the sample. The calorimetry study is complemented by resistivity measurements that give extra information about the superconducting and normal states.

In figure 1a we show the specific heat of a large crystal (8mg) measured at ambient pressure compared to the measurement on the small crystal in the diamond anvil cell at zero pressure. For the large sample the superconducting transition appears as a sharp step at 1.45K. The sample is therefore homogeneous even though the critical temperature is slightly lower than some reported values. The low temperature side shows that $C/T$ extrapolates to a residual value of about 60-70 mJ/mol.K$^2$ as is found systematically in UTe$_2$. As we mentioned already, the measurement in the diamond anvil cell is not quantitative. Still the transition shows up as a sharp anomaly with a similar aspect to the ambient pressure measurement. It is slightly broadened either due to some very small pressure, to some degradation of the sample during the set up of the cell, or to the measurement technique. The transition appears on top of a temperature dependent background. Figure 1, graphs b, c and d show the evolution of the temperature dependence of $C/T$ for 3 pressure ranges. At low pressure, the anomaly at $T_{S1}$ corresponding to the superconducting transition clearly shifts to lower temperatures. Above 0.3 GPa, a second anomaly appears at a higher temperature, labeled $T_{S2}$. $T_{S2}$ then continues to increase with pressure. Figure 1c shows the intermediate pressure range: $T_{S1}$ continues to decrease and falls below the lowest measurable temperature (about 0.5K) above 0.8 GPa. Above 1 GPa the anomaly corresponding to $T_{S2}$ becomes much more pronounced. $T_{S2}$ continues to increase, reaching a maximum of about 2.8 K at 1.3 GPa, then decreases rapidly and the anomaly disappears above 1.5 GPa. However as shown in figure 1d, as pressure is further increased above 1.8 GPa, a new anomaly appears at a temperature labeled $T_{M3}$, of about 3.5 K. This last anomaly, initially weak, becomes more pronounced with pressure and moves to higher temperatures.

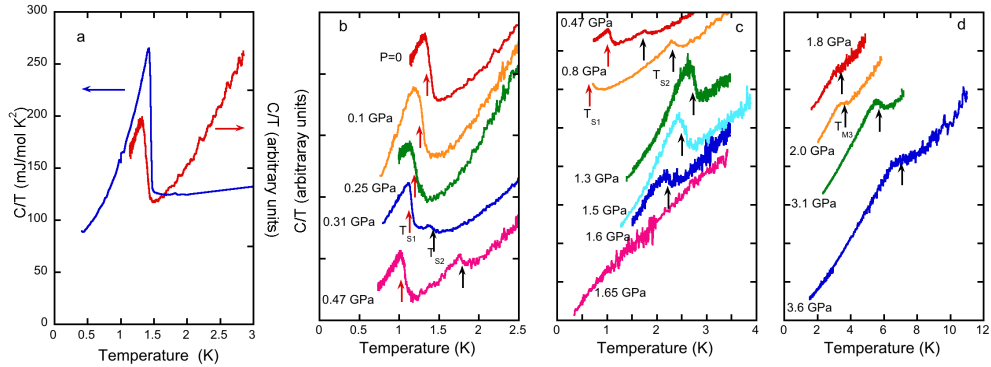

*Figure 1 : Graph a shows the specific heat of a large sample from the same batch measured at ambient pressure and the measurement at zero pressure in the diamond anvil cell. Graphs b, c and d show measurements under pressure in different pressure ranges. The curves have been shifted vertically for clarity.*

Specific heat alone is not sufficient to identify the new phases that appear. By continuity there seems little doubt that $T_{S1}$ corresponds to the superconducting phase seen at ambient pressure. We will show that the effect of magnetic field on this transition is also consistent with ambient pressure measurements. In order to gain information on $T_{S2}$ and $T_{M3}$ we compare with resistivity measurements. Figure 2 shows $\rho(T)$ curves for different pressures. The superconducting critical temperature increases with pressure reaching a maximum of about 2.7K between 0.5 and 1 GPa. This implies that the anomaly seen at $T_{S2}$ in the specific heat is also due to superconductivity. On increasing pressure the superconducting critical temperature in the resistivity experiment starts to decrease, consistent with the specific heat results.

At higher pressure no superconductivity is seen, and a new anomaly appears. This corresponds to the temperature $T_{M3}$ in the specific heat measurement and shows up as a small increase in resistivity as the temperature decreases. The most likely explanation is the appearance of magnetic order, although this remains to be checked by other types of measurements. The phase diagrams obtained from resistivity and specific heat are compared in figure 4. The spacing of the pressure points for the resistivity experiment was too large to resolve the initial decrease of the critical temperature. Also the calorimetry experiment finds a strong increase in $T_{M3}$ at high pressure, which was not observed in the resistivity, possibly because the maximum pressure was not high enough. There is also a slight difference in the pressure scale between the two experiments, probably due to the different pressure conditions (see methods), but the two phase diagrams are consistent.

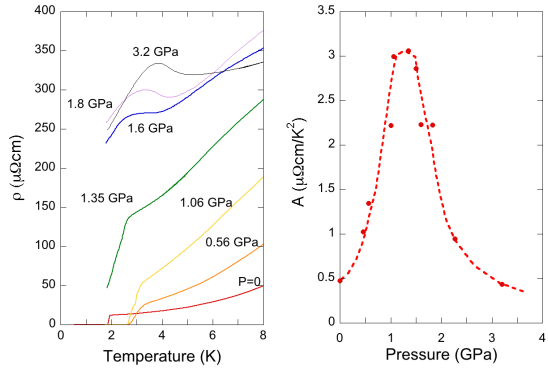

*Figure 2 : Left graph shows resistivity curves for different pressures. Right graph, the pressure evolution of the pre-factor A obtained by fitting the resistivity temperature dependence with a Fermi liquid law of the form $\rho= \rho_0+AT^2$. The dashed line is a guide for the eye.*

More information can be gained from the normal state temperature dependence of the resistivity which can be fitted with a Fermi liquid behavior of the form $\rho= \rho_0+AT^2$. In the right-hand graph of figure 2, we show the pressure dependence of the pre-factor $A$. The value of $A$ can be approximately related to the electronic effective mass, usually obeying an empirical relationship $A^{1/2}\approx m^*$[13], although this relation may break down close to an instability. $A$ increases quite significantly with pressure by a factor 6 between 0 and 1 GPa, and has a maximum around 1.3 GPa consistent with the existence of a critical pressure in this region. At higher pressures $A$ decreases although its determination is less reliable because of the influence of the magnetic transition meaning that $A$ has to be determined at higher temperatures. We note that this enhancement of a factor 6 of $A$ is quite similar to that found under high field at the metamagnetic transition[7]. The evolution of $m^*$ with field has also been estimated from magnetization and magnetocaloric measurements and an enhancement of a factor 1.5 – 2 is found at the metamagnetic transition[10,14]. Interestingly it was shown that this enhancement of $m^*$, if due to an increase of the pairing strength, would imply an increase in the superconducting critical temperature to about 3K[10,14], quite similar to the value we find under pressure.

The effect of an applied field is also instructive. We could apply a field only along the c-axis. In figure 3 we show the field dependence of the different transitions at several pressures. The slope of the upper critical field $H_{C2}$ at zero pressure is consistent with previous reports[5,6]. At low pressure this slope decreases slightly, scaling with $T_{S1}$ as expected. When the second transition appears at $T_{S2}$, its field dependence is initially quite similar to that of $T_{S1}$. However as pressure is increased the slope becomes extremely steep. The resistivity measurement also reveals a large increase in the slope. In a simple picture the slope should scale with both the critical temperature and the square of the effective mass. The first term gives a factor of 1.8. Using the relation $m^*\approx A^{1/2}$ would imply an enhancement of a factor 6 for $m^{*2}$. This is the right order of magnitude to explain the increase by more than a factor 10 of the slope of $H_{C2}$, however this method probably largely overestimates the enhancement of m*. Another approach is to reproduce the enhancement of $T_{s2}$ between 0.3 and 1.3 GPa by an increase of the strong coupling constant $\lambda$. The expected slope can then be calculated taking into account the renormalization of $m^*$ due to the increase of $\lambda$. This accounts quite well for the increase of the slope of $H_{C2}$ up to 1 GPa, however it only accounts for about 50% of the maximum measured slope of over 100T/K (see supplemental Material for more details).

.

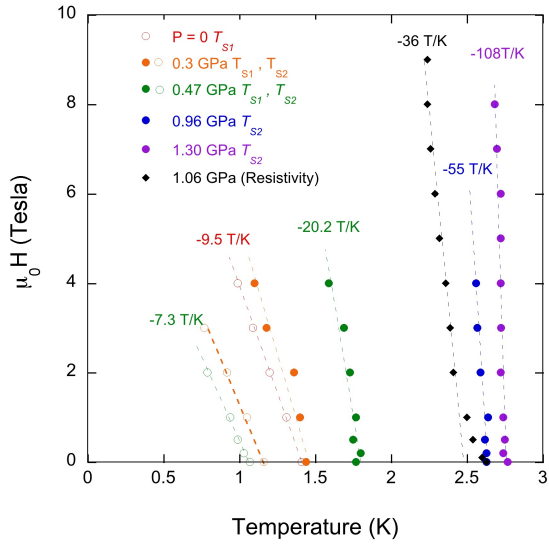

*Figure 3 : Field dependences of the 2 transitions for different pressures. The field dependence of the superconducting transition measured by resistivity at 1.06 GPa is also included for comparison.*

In figure 4 we show the general phase diagram obtained from the specific heat measurements. The strong enhancement of the critical temperature, with $T_{S2}$ reaching 2.8 K makes UTe$_2$ the heavy fermion superconductor with the highest critical temperature among all Ce- and U-based systems known so far. Higher values have only been found so far in neptunium and plutonium based heavy fermion compounds[15,16]. The most striking result is the discovery of two superconducting states with different transition temperatures. As stated in the introduction this is an extremely rare phenomenon. Both transitions are visible in the specific heat implying that they are bulk phases. Pressure has opposite effects on the two transition temperatures, $T_{S1}$ decreases while $T_{S2}$ increases with increasing pressure. The effect of field applied along the $c$-axis on the two transitions, while initially similar, becomes also quite different at high pressure. These different pressure and field effects are strong arguments for the two phases being of quite different nature. In general, multiple superconducting transitions are expected from either lifting the degeneracy of a multicomponent order parameter, or from two different irreducible representations of the space group that are accidentally nearly degenerate. In UPt$_3$ the consensus favors nowadays the former[17-20]. However, as UTe$_2$ is orthorhombic, all irreducible representations are one-dimensional so only the second scenario is allowed. This however means that the two transitions should cross rather than split, as suggested by the dotted line in figure 4 even though within the experimental sensitivity no indication of $T_{S2}$ lying below $T_{S1}$ is seen at low pressure. In this, UTe$_2$ resembles more the case of UBe$_{13}$, where a small amount of doping with thorium apparently splits the single superconducting transition into two separate transitions[21,22], but in fact an observed change in the order parameter implies the existence of a fourth line in the phase diagram even though no transition has been directly detected[21,23,24]. We have sketched a similar line with a question mark in figure 4.

There is another intriguing possibility that allows only two superconducting phases. Generally, a phase diagram with three second-order transition lines meeting at a single multicritical point is thermodynamically forbidden except for the very special case, where the specific heat jump on one of the lines vanishes at the crossing point and the slopes of the other two lines are identical, (see more details in Methods and Supplemental Material). In fact, in our results the jump in $C/T$ at $T_{S2}$ is initially very small, and the slope of $T_{S1}$ is identical on both sides of the meeting point, so possibly the solution with just three phase boundaries could be thermodynamically allowed here. Nevertheless the small initial amplitude in the jump in $C/T$ at $T_{S2}$ is puzzling, and seemingly difficult to reconcile with a bulk transition between normal and superconducting states. This is in contrast to the cases of $UBe_{13}$ or $UPt_3$, where the two specific heat jumps have similar amplitudes. The steep slope of $H_{C2}$ implies that the effective mass of the electrons condensing at $T_{S2}$ is large, and so should contribute to the specific heat. We cannot exclude that for some reason only part of the sample volume is initially superconducting at $T_{S2}$, but the fact that the transition is detected in the specific heat means that this fraction is non-negligible. In fact at higher pressure the jump in $C/T$ at $T_{S2}$ becomes larger than that for $T_{S1}$ at ambient pressure, and taking into account the probable increase in $m^*$ their amplitudes are quite similar (see supplemental material). Other results also point to the phase at $T_{S2}$ being an intrinsic property of $UTe_2$: The normal state resistivity shows that the appearance of the high pressure phase at $T_{M3}$ and the strong increase of $A$ are fully bulk effects, and the superconducting phase at $T_{S2}$ is related to both of these, disappearing at the onset of the high pressure phase, and with the maximum of $T_{S2}$ being associated with the maximum of $A$. The small jump in $C/T$. may be related to the still unsolved question of the residual specific heat at ambient pressure which has been attributed to a non-unitary superconducting state, with only half of the electrons condensing [5]. While this point still has to be established, a possible explanation for our result could be that under pressure at $T_{S2}$, an even smaller fraction of the electrons initially condenses, concerning just a small part of the Fermi surface, which progressively increases with pressure. An almost gapless order parameter could also be responsible for a reduced jump in $C/T$. What we miss at this stage is a complete entropy balance, which cannot be estimated because the ac calorimetry technique used here is not quantitative, and the measurements do not extend to low enough temperatures.

The other main point is that pressure clearly drives $UTe_2$ towards a more strongly correlated state. The strong increase of the $A$ coefficient implies a large enhancement of the effective mass. At the same time superconductivity becomes much more robust to a magnetic field applied along the $c$-axis, more than expected from the increase in the strong coupling constant: the most likely explanation for this anomalous slope is that the magnetic field actually enhances the pairing strength as seen in the ferromagnetic superconductors[25,26], and in $UTe_2$ at zero pressure. So far a spectacular reinforcement of superconductivity has been found in $UTe_2$ for a field applied along the b-axis, and also at a specific angle in the b-c plane[5,8,9]. Our results suggest that under pressure field-reinforced superconductivity probably also exists for a field along the c-axis. At a pressure of 1.7 GPa no clear anomaly corresponding to any transition is seen down to about 0.5K, the lowest temperature measured, however at 1.8 GPa a new phase transition is observed at about 3.5 K, probably due to a magnetically ordered state. This is seen in both the specific heat and in the resistivity. The resistivity shows a small but clear increase as temperature decreases through this phase transition. Such a phenomenon is usually associated with a change in the Fermi surface due to a change in the periodicity of the unit cell. It is therefore unlikely that the ordered state is a simple ferromagnetic order. We can speculate that it if some modulated structure develops, it could be responsible for the suppression of ferromagnetic fluctuations, and of superconductivity.

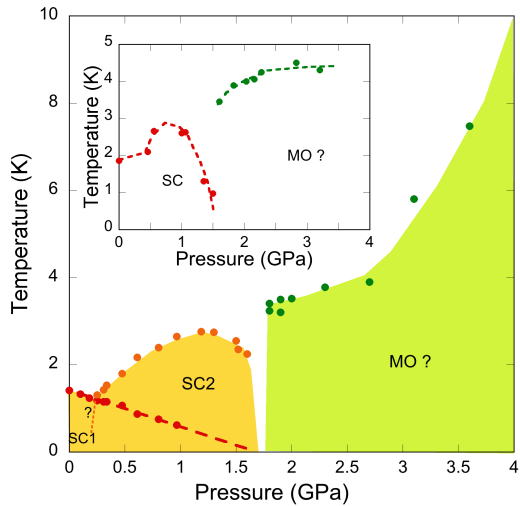

UTe$_2$ is a fascinating system that challenges many of our ideas about heavy fermion superconductivity. Clearly its high-pressure properties are no exception. UTe$_2$ under pressure combines two of the most intriguing effects found in unconventional superconductors, namely multiple superconducting order parameters and field reinforced superconductivity. These are associated with the highest known superconducting critical temperature for a Ce- or U- based heavy fermion system. The similarities in the reinforcement of correlations and superconductivity with pressure and with high magnetic field are quite striking. Under pressure we find a field reinforcement of superconductivity for the field applied along the *c*-axis whereas at ambient pressure this effect was only seen for field applied closer to the *b*-axis. This suggests a change in the magnetic and/or electronic anisotropies and we can expect spectacular effects under combined pressure and very high fields. Further studies to obtain a more complete picture are now under way and will no doubt improve our understanding of these phenomena and perhaps bring more exciting discoveries.

Methods
Single crystals of UTe$_2$ were grown by vapor transport as described in more detail here[6]. The ambient pressure specific heat was measured in a commercial device (QD PPMS). The ac calorimetry was measured in a diamond anvil cell as described here[27]. The transmitting medium was argon ensuring very hydrostatic conditions in this pressure range. The sample was heated by a laser diode at a frequency of 637 Hz and its temperature oscillations were measured with a Au/Au:Fe thermocouple. The pressure was tuned in-situ in a dilution refrigerator and measured with the ruby luminescence scale. Because of the heat load from the laser diode the lowest temperature achievable was about 0.5 K. The resistivity was measured in a modified Bridgman cell with ceramic anvils as described here[28]. The pressure-transmitting medium was Fluorinert. The sample for resistivity came from a different batch and showed a higher transition temperature. Measurements were performed down to 1.8K in a commercial device (Quantum Design PPMS). In both experiments a magnetic field up to 8 or 9 T could be applied along the c-axis of the crystal. The theoretical analysis of thermodynamically allowed phase diagrams is performed by the Taylor expansion of the Gibbs free energy in the vicinity of the multicritical point in the *P-T* diagram. The evaluation of the pressure evolution of the slopes with that of the critical temperatures uses the same strong coupling model as used for UCoGe[26]. Further details are provided in the Supplemental Material.

Acknowledgments
We acknowledge the financial support of the Cross-Disciplinary Program on Instrumentation and Detection of CEA, the French Alternative Energies and Atomic Energy Commission.

Author contributions
D.B. supervised the project and performed the calorimetry measurements with F.H. M.V. performed the resistivity measurements with G.K. G.L. grew the single crystals with input from D.A. D.B. and M.V. analyzed the data. J.-P.B. performed the strong coupling analysis. M.E.Z. performed the thermodynamic analysis. D.B. wrote the manuscript with input from M.V.,J.F., J-P.B., A.P., M.E.Z., G.K. and D.A.

# Supplemental material to

# Multiple Superconducting Phases in a Nearly Ferromagnetic System

D. Braithwaite,[1] M. Valiska,[1] G. Knebel,[1] G. Lapertot,[1] J.-P. Brison,[1]
A. Pourret,[1] M. E. Zhitomirsky,[1] J. Flouquet,[1] F. Honda,[2] D. Aoki[2]

[1] *Univ. Grenoble Alpes, CEA, IRIG, PHELIQS, 38000 Grenoble, France,*
[2] *IMR, Tohoku University, Oarai, Ibaraki, 311-1313, Japan*

### A. Determination of $H_{c2}$ from calorimetry

In Figure S1 we show $C/T$ plots under applied magnetic field for two pressures. At low pressure the transition at $T_{S1}$ remained sharp and could be easily followed down to the lowest temperatures measured. The transition at $T_{S2}$ quickly broadened making a reliable determination of the phase diagram difficult even up to 4 T. However at higher pressure the transition at $T_{S2}$ remained extremely sharp even at the maximum field of 8 T.

### B. Determination of the $A$ coefficient

For low pressures the temperature range used was the lowest range above the superconducting transition where a linear dependence in the plot of $\rho$ versus $T^2$ was found. At higher pressure the fit was made avoiding the beginning of the upturn due to $T_{M3}$, meaning the fit was typically made in the temperature range 6–8 K (see Figure S2).

### C. Estimation of the slope of $H_{c2}$ from a strong coupling model

The slope of the upper critical field at at $T = T_s$ is sensitive both to the superconducting transition temperature $T_s$ and to an average Fermi velocity $v_F$ perpendicular to the applied field. In most cases, the pressure dependence of $T_s$ is governed by the pressure dependence of the strong coupling constant $\lambda$, which is the most sensitive parameter (in the weak-coupling regime, $T_s$ depends exponentially on $\lambda$). In turn, the pressure dependence of $\lambda$ implies a pressure dependence of $v_F$, through the Eliashberg relation:

$$v_F = \frac{v_F^{\text{band}}}{1+\lambda} \ ,$$

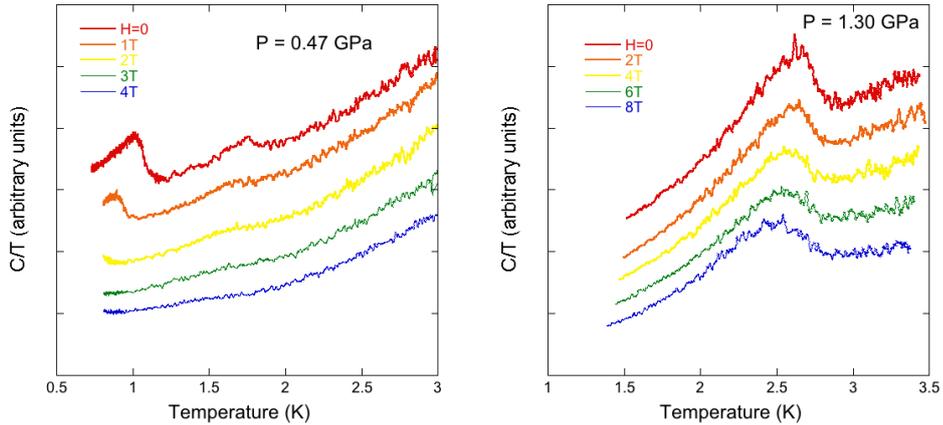

Fig. S 1: $C/T$ curves for different fields for pressures of 0.47 GPa (left) and 1.30 GPa (right).

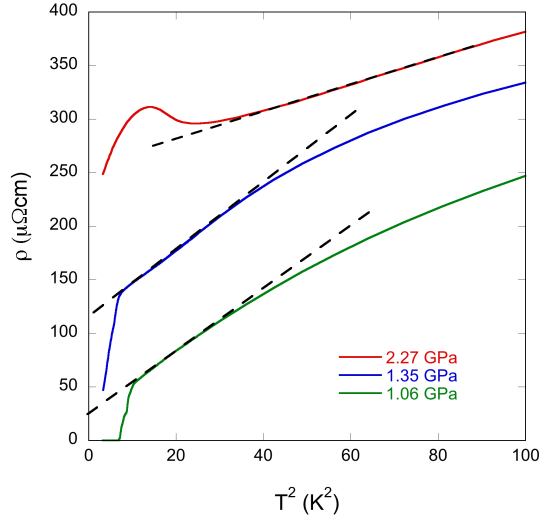

Fig. S 2: Resistivity curves versus $T^2$ showing the typical temperature ranges used to extract the $A$ coefficient from a Fermi liquid behaviour $\rho = \rho_0 + AT^2$.

where $v_F^{\text{band}}$ is the band Fermi velocity renormalized by all interactions but the pairing mechanism. Using a minimal strong coupling model [1] and starting from a strong coupling constant at zero pressure of $\lambda = 1.5$, which gives the right order of magnitude for the observed change of the specific heat under field along the $b$-axis [2, 3], we obtain the pressure variations of $\lambda$ reproducing the evolution of $T_{s1}$ and $T_{s2}$ shown in Figure S3. Note that above 0.3 GPa, the transition at $T_{s1}$ is most likely a transition between two superconducting phases, so that the calculation, strictly speaking, does not apply anymore. With the same model, we have adjusted $v_F$ at $P = 0$ to reproduce the experimental slopes of $H_{c2}$ for the transition at $T_{s1}$, and at $P = 0.3$ GPa for the transition at $T_{s2}$, and then calculated how they would change following the sole evolution of $\lambda$ (see Fig. S3). The strong enhancement of the $\partial H_{c2}/\partial T$ at $T_{s2}$ is well reproduced up to 1 GPa, but clearly, an additional mechanism is required to understand the huge slope of $H_{c2}$ at 1.7 GPa. Again, above 0.3 GPa, if the transition at $T_{s1}$ is a transition between two superconducting phases, the calculation of the slope is not relevant, and it is only shown as a reference.

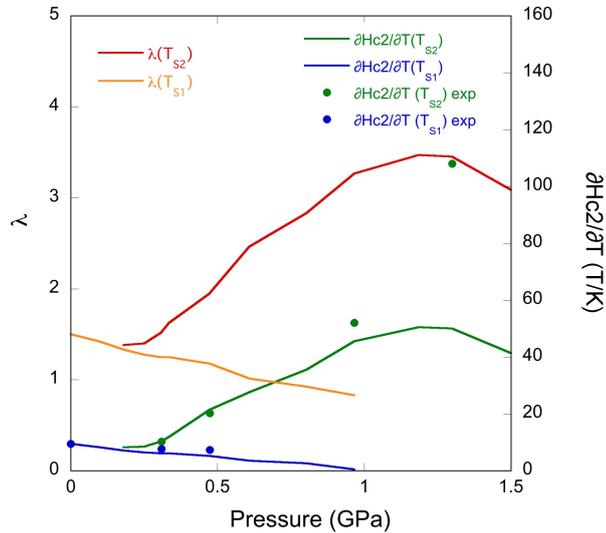

Fig. S 3: Estimation (left scale) of the evolution of $\lambda_1$ and $\lambda_2$ from the change in $T_{S1}$ and $T_{S2}$ and (right scale) corresponding expected values of the slope of $H_{c2}$. The correspondence is good up to 1 GPa, but at 1.3 GPa the experimental slope is twice larger than expected, implying some other mechanism, probably a further enhancement of the pairing mechanism with field.

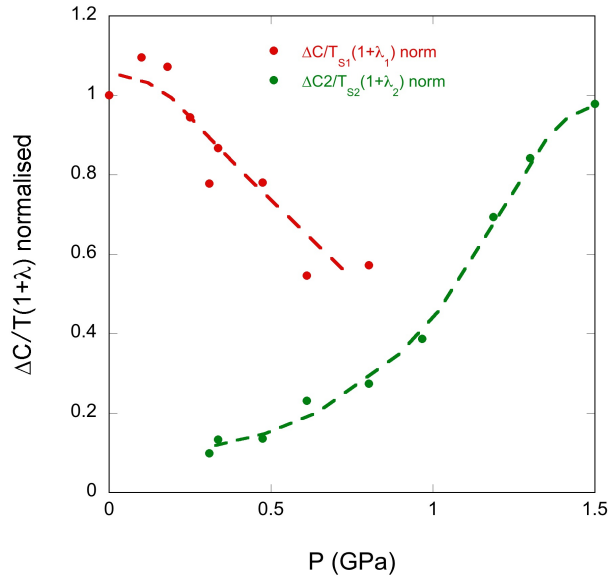

Fig. S 4: Amplitude of the $C/T$ jumps at $T_{S1}$ and $T_{S2}$, multiplied by $(1+\lambda)$ assuming $m^* \approx (1+\lambda)$ and normalized to the value of the jump at $T_{S1}$ at ambient pressure. Once more, if the transition at $T_{S1}$ is between two superconducting transitions, the normalization for $\Delta C/T_{s1}$ above 0.3 GPa is not correct.

### D. Semi-quantitative analysis of the jumps $C/T$ at the two transitions

Although the ac calorimetry measurement is not quantitative, we can compare amplitudes of the specific heat jumps $\Delta C/T$ in relation to each other at different pressures, see Figure S4. Still, caution should be used as the evolution of their amplitude with pressure is influenced by various factors. Nevertheless, it is clear that the jump at $T_{S2}$ is progressively reinforced in relation to that at $T_{S1}$ with pressure. The amplitude of the jump at $T_{S2}$ at high pressure actually seems to become larger than that at $T_{S1}$ at ambient pressure. However taking into account the enhancement of $m^*$ due to the increase of $\lambda$ (deduced from the previous calculation, see Figure S3), assuming $m^* \approx (1+\lambda)$, these amplitudes are probably quite similar.

### E. Thermodynamic considerations on the phase diagram

Our aim is to verify thermodynamic conditions for the appearance of a phase diagram with the topology sketched in Figure S5. Three different phases labeled $A$, $B$, and $N$ meet at a multicritical point $M = (T^*, P^*)$ such that three boundary lines $AN$, $BN$, and $AB$ are all lines of continuos (second order) transitions. We do not make any assignment on the nature and broken symmetries in each of the phases. Note, that the standard textbook description of possible phase diagrams, see, for example, [4], includes only a diagram with the discontinuous first-order transition line between phases $A$ and $B$. Another type of phase diagrams commonly appearing in the Landau theory of phase transitions is a diagram with *four* lines of continuous transitions meeting at a single tetracritical point.

In our analysis, we follow the approach used by Leggett in the analysis of the phase diagram of superfluid $^3$He [5], see also application to unconventional superconductors [6]. We begin by noting that the entropy $S_X(P, T)$ and the volume $V_X(P, T)$ of each of the three phases ($X = A, B, N$) tend to the same common values $(S^*, V^*)$ as one approaches the multicritical point $M = (P^*, T^*)$. In such a case the Gibbs potentials $G_X$ for each of the phases have equal linear terms in the expansion in $\Delta T$, $\Delta P$ near the $M$ point. To study the phase coexistence one has to expand $G_X$ up to the second order:

$$\Delta G_X = -\frac{1}{2}\,\alpha_X(\Delta T)^2 + \beta_X\,\Delta T\Delta P - \frac{1}{2}\,\gamma_X(\Delta P)^2 \;, \tag{1}$$

where second order derivatives of the Gibbs energy are

$$\alpha = \left(\frac{\partial S}{\partial T}\right)_P \;, \qquad \beta = \left(\frac{\partial V}{\partial T}\right)_P = -\left(\frac{\partial S}{\partial P}\right)_T \;, \qquad \gamma = -\left(\frac{\partial V}{\partial P}\right)_T \;. \tag{2}$$

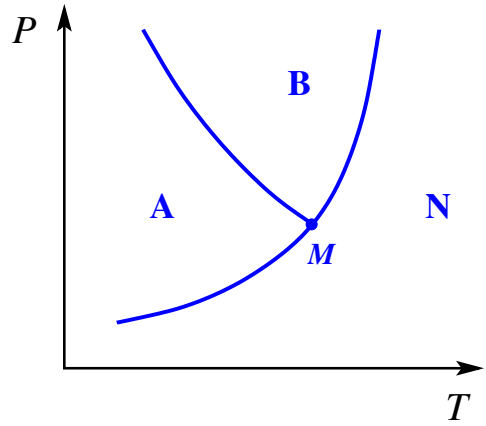

Fig. S 5: The $P$–$T$ diagram with three lines of second-order transitions meeting at a multicritical point $M$.

Coefficients $\alpha = C/T$ and $\gamma = V\kappa_T$ are proportional to the specific heat $C$ and the isothermal compressibility $\kappa_T$ and, hence, are positive.

We denote slopes of the phase boundaries in the vicinity of the $M$ point by

$$r_1 = \left(\frac{dP}{dT}\right)_{AN} , \qquad r_2 = \left(\frac{dP}{dT}\right)_{BN} , \qquad r_3 = \left(\frac{dP}{dT}\right)_{AB} . \tag{3}$$

According to the Ehrenfest relations [4], these are expressed via the jumps of the second-order derivatives in the thermodynamic potential

$$r_1 = \left(\frac{\Delta\alpha_{AN}}{\Delta\beta_{AN}}\right) = \left(\frac{\Delta\beta_{AN}}{\Delta\gamma_{AN}}\right) , \quad r_2 = \left(\frac{\Delta\alpha_{BN}}{\Delta\beta_{BN}}\right) = \left(\frac{\Delta\beta_{BN}}{\Delta\gamma_{BN}}\right) , \quad r_3 = \left(\frac{\Delta\alpha_{AB}}{\Delta\beta_{AB}}\right) = \left(\frac{\Delta\beta_{AB}}{\Delta\gamma_{AB}}\right) . \tag{4}$$

From the first two relations we express $\Delta\beta$ and $\Delta\gamma$ on the first two boundaries via $\Delta\alpha$ and the corresponding slope and substitute those into the expression for $r_3$. As a result we obtain

$$(\Delta\alpha_{AN} - \Delta\alpha_{BN})\left(\frac{\Delta\alpha_{AN}}{r_1^2} - \frac{\Delta\alpha_{BN}}{r_2^2}\right) = \left(\frac{\Delta\alpha_{AN}}{r_1} - \frac{\Delta\alpha_{BN}}{r_2}\right)^2 . \tag{5}$$

The above equation is further simplified to

$$\Delta\alpha_{AN}\Delta\alpha_{BN}\left(\frac{1}{r_1} - \frac{1}{r_2}\right)^2 = 0 . \tag{6}$$

Thus, in order to allow crossing of three lines of second-order transitions in a single point one requires either equal slopes $r_1 = r_2$ or vanishing specific heat jump $\Delta\alpha = 0$ on one of the boundaries. Applying Eq. (6) subsequently to three boundaries, we obtain that two transition lines with finite jumps $\Delta\alpha$ ($\Delta C$) have equal slopes, whereas the third boundary exhibits the specific heat anomaly that vanishes towards the multicritical point.

In the end, we would like to remark that the thermodynamic stability of the discussed phase diagram does not necessarily mean that there is a simple phenomenological Landau functional, which may produce such a diagram. Large variation of the specific heat jump $\Delta C/T$ on boundary of $T_{S2}(P)$ provide strong constraints on future microscopic theories of superconducting phase coexistence in UTe$_2$.

--------